\documentclass[prd,aps,showpacs,nofootinbib,eqsecnum,onecolumn,groupedaddress,amssymb]{revtex4}
\usepackage{amsmath}
\usepackage{amssymb}
\usepackage{amsfonts}
\usepackage{graphicx,bm}
\usepackage{dcolumn}
\usepackage{color,amsxtra}
\usepackage{epsf}
\usepackage{enumerate}
\usepackage{hhline}
\usepackage{array}
\usepackage{tabularx}
\usepackage{subfigure}
\usepackage{fancyhdr}
\usepackage{mathrsfs}

\newcommand{\be}{\begin{equation}}
\newcommand{\ee}{\end{equation}}
\newcommand{\bea}{\begin{eqnarray}}
\newcommand{\eea}{\end{eqnarray}}
\newcommand{\beaa}{\begin{eqnarray*}}
\newcommand{\eeaa}{\end{eqnarray*}}

\newcommand{\Eqn}[1]{&\hspace{-0.2em}#1\hspace{-0.2em}&}

\def\be{\begin{equation}}
\def\ee{\end{equation}}
\def\bea{\begin{eqnarray}}
\def\eea{\end{eqnarray}}

\begin{document}

\title{Trace-anomaly driven inflation in modified gravity and the BICEP2 result}
\author{Kazuharu Bamba$^{1, 2, 3}$, 
R.~Myrzakulov$^{4}$, 
S.~D.~Odintsov$^{5, 6, 7, 8}$
and
L.~Sebastiani$^{4}$
}
\affiliation{
$^1$Leading Graduate School Promotion Center,
Ochanomizu University, 2-1-1 Ohtsuka, Bunkyo-ku, Tokyo 112-8610, Japan\\
$^2$Department of Physics, Graduate School of Humanities and Sciences, Ochanomizu University, Tokyo 112-8610, Japan\\
$^3$Kobayashi-Maskawa Institute for the Origin of Particles and the
Universe, Nagoya University, Nagoya 464-8602, Japan\\
$^4$Department of General \& Theoretical Physics and Eurasian Center for Theoretical Physics, Eurasian National University, Astana 010008, Kazakhstan\\ 
$^5$Consejo Superior de Investigaciones Cient\'{\i}ficas, ICE/CSIC-IEEC, 
Campus UAB, Facultat de Ci\`{e}ncies, Torre C5-Parell-2a pl, E-08193
Bellaterra (Barcelona), Spain\\
$^6$Instituci\'{o} Catalana de Recerca i Estudis Avan\c{c}ats
(ICREA), Barcelona, Spain\\ 
$^7$Tomsk State Pedagogical University, 634061 Tomsk and 
National Research Tomsk State University, 634050 Tomsk, Russia\\
$^8$King Abdulaziz University, Jeddah, Saudi Arabia
}


\begin{abstract}
We explore conformal-anomaly driven inflation in $F(R)$ gravity without invoking the scalar-tensor representation. 
We derive the stress-energy tensor of the quantum anomaly in the flat 
homogeneous and isotropic universe.  
We investigate a suitable toy model of exponential gravity 
plus the quantum contribution due to the conformal anomaly, which 
leads to the de Sitter solution. 
It is shown that in $F(R)$ gravity model, 
the curvature perturbations with its enough amplitude consistent with 
the observations are generated during inflation. 
We also evaluate the number of $e$-folds at the inflationary stage 
and the spectral index $n_\mathrm{s}$ of scalar modes of the curvature perturbations by analogy with scalar tensor theories, and compare them 
with the observational data. 
As a result, it is found that the Ricci scalar decreases during inflation and 
the standard evolution history of the universe is recovered at the small curvature regime. Furthermore, it is demonstrated that in our model, 
the tensor-to-scalar ratio of the curvature perturbations can be a finite value within the $68\%\,\mathrm{CL}$ error of the very recent result found by the BICEP2 experiment. 
\end{abstract}

\hspace{13.1cm} OCHA-PP-319

\maketitle

\def\thesection{\Roman{section}}
\def\theequation{\Roman{section}.\arabic{equation}}

\section{Introduction}

It is theoretically known that the de Sitter expansion in the early universe, 
i.e., inflation~\cite{Guth, Sato}, can be realized by the quantum effects of conformally invariant fields~\cite{DC-MM} (for recent reviews on inflation, 
see, e.g.,~\cite{revinflazione, rr2, Linde}). 
This has also been expected by the effect of quantum theories in vacuum 
and that of quantum gravity~\cite{BOS, MW}. 
Inflation is a crucial process to solve the so-called horizon and flatness problems, and to produce the curvature perturbations, which are the seeds of the large-scale structure of the universe. 

For the standard inflation models such as the chaotic inflaion~\cite{chaotic}, 
inflation is driven by the so-called inflaton field. 
If the inflaton potential has a minimum, the inflaton rolls down to it 
and oscillates, so that the reheating process can occur. 
There are also the other models in which the inflaton falls 
in a potential hole~\cite{buca1, buca2, buca3, buca4}, 
or the hybrid or double inflation~\cite{ibrida1, ibrida2}, 
where inflation is realized by two scalar fields appearing 
through a phase transition. 
However, the mechanism for the initial confinement of the inflaton 
has not well understood yet. 
There have been proposed a model of trace-anomaly driven inflation, 
called Starobinsky inflation~\cite{Staro}, 
and its extended version~\cite{Alex} 
by examining the coefficient of the trace-anomaly higher-(total-)derivative 
counter term more carefully. 

On the other hand, it is observationally suggested that 
the trace-anomaly driven inflation or 
its similar version with an $R^2$ term 
could be one of the most realistic candidates of inflationary theory 
by recent cosmological observations~\cite{WMAP, 
Hinshaw:2012aka, Ade:2013lta, Ade:2013uln}. 
Accordingly, various aspects of an 
$R^2$ inflation model have been investigated in the literature 
(for related works, see, for instance,~\cite{R2inf, Huang:2013hsb, FRinfl}). 
Inflation with enough duration and its graceful exist can occur 
due to the trace-anomaly $\Box R$ term~\cite{Staro}, 
which is a higher-derivative $R^2$ term in the anomaly-induced action. 
In addition, this is the effective action of gravity including an $R^2$ term 
and a non-local one. Hence, thanks to these terms, it can be interpreted as a modified gravity theory. 
Also, the trace-anomaly driven inflation has been examined in several extended gravity theories such as 
gravity with anti-symmetric torsion~\cite{Buc}, 
dilaton-coupled gravity~\cite{GOZ-NO}, 
where in the conformal anomaly, there exist 
additional terms with their dilaton dependence, 
and gravity in the so-called brane new world~\cite{HHR-NOZ-NO-CI, Haw}. 

In this paper, following the considerations in Ref.~\cite{OdCA}, 
we explore inflation due to the quantum anomaly in $F(R)$ gravity 
(for recent reviews, see, for example,~Ref.~\cite{R-NO-CF-CD} on modified gravity and Ref.~\cite{Bamba:2012cp} on dark energy). 
If inflation happens during the Planck epoch, the quantum effects given by the trace anomaly of a large number of matter fields have to be taken into account. 
We investigate the effects of a modification of gravity 
on the primordial de Sitter universe emerging due to the conformal anomaly, the generation of the curvature perturbations, and the graceful exit from inflation occurs. 
To carry out our analysis, we use a simple toy model of exponential gravity 
involving the quantum contribution owing to the conformal anomaly, in which the late-time cosmic acceleration can occur in a unified way with inflation in the early universe. 
This model is adopted for our purpose, because it makes possible the de Sitter expansion driven by conformal anomaly involving negative perturbations on the curvature which starts to decrease towards the end of inflation. 
We calculate the number of $e$-folds $\mathcal N$ 
and the spectral index of spectrum of the curvature perturbations generated 
during inflation in analogy with the scalar field theories, and compare them with the cosmological observations. 
We demonstrate that a viable inflation with the end in the Planck epoch can be realized. 
In this approach, the role of the inflaton is played by the modification of gravity, and instead to its decay in the scalar and the Dirac fields during the reheating, the model turns out to be the Einstein gravity at small curvature 
at the radiation-dominated stage with the decelerated expansion of 
the universe. 
We use units of $k_{\mathrm{B}} = c = \hbar = 1$ and denote the 
gravitational constant, $G$, by $\kappa^2\equiv 8 \pi G$, so that
$G=1/M_{\mathrm{Pl}}^2$ with $M_{\mathrm{Pl}} =1.2 \times 10^{19}$ GeV the Planck mass. The metric signature $(-, +, +,+)$ is adopted.

The paper is organized as follows. 
In Sec.~II, 
we review the trace anomaly from quantum corrections in the Yang-Mills theory coupled with gravity. The introduction of an $R^2$ term in the action of the Einstein gravity is discussed. 
In Sec.~III, 
the trace anomaly is explored in $F(R)$ gravity and the explicit form of the stress-energy tensor of quantum corrections for the flat 
Friedmann-Lema\^{i}tre-Robertson-Walker (FLRW) space-time is derived. 
Section IV is devoted to the study of the trace-anomaly driven inflation in exponential gravity. The (unstable) de Sitter solution describing the early-time acceleration will be found and the dynamics of inflation will be investigated. The explicit form of the perturbations coming from the modification of gravity will be derived showing the mechanism which makes possible a graceful exit from the inflation. We pay attention to the evaluation of the number of $e$-folds $\mathcal N$, the spectral index of scalar modes of the curvature perturbations, 
and some explicit example of the parameterization of the model is presented. 
In particular, we explicitly analyze 
the spectral index of scalar modes of the curvature perturbations and 
those tensor-to-scalar ratio and compare them with the recent observational data obtained by the Planck~\cite{Ade:2013lta, Ade:2013uln} satellite 
and the BICEP2~\cite{Ade:2014xna} experiment. 
In the last part of this Section, we examine the possibility to recover the standard thermal universe at small curvature. 
In Sec.~V, we study a unified scenario between trace-anomaly driven inflation and the late-time cosmic acceleration in exponential gravity.
Conclusions, discussions and final remarks are presented in Sec.~VI.

\section{Trace anomaly}

Since the Standard Model of particle physics contains almost a hundred fields and this number may further be doubled if the Standard Model is embedded in a supersymmetric theory, it is reasonable to consider that there exist a large number of matter fields during inflation in the early universe.  
The action of these (massless) matter fields (scalars, the Dirac spinors, and vectors) in curved space-time is conformal invariant, 
but some divergences appear due to the presence of 
the one-loop vacuum contributions. 
In the renormalized action, some counterterms are called to cancel the poles of the divergence part by the price of breaking the conformal invariance of the matter action itself. {}From the classical point of view, 
the trace of the energy momentum tensor in a conformally invariant theory is null. However, the renormalization procedure leads to the trace of 
an anomalous energy momentum tensor, the so-called quantum anomaly. 
In the four-dimensional space-time, it reads~\cite{Deser, Birel, Duff}
\begin{equation}
\langle T_{\mu}^\mu\rangle=\alpha\left(W+\frac{2}{3}\Box R\right)-\beta \mathcal{G}+\xi\Box R\,,
\label{TA}
\end{equation}
where $\langle T_{\mu}^\mu\rangle$ is the vacuum expectation value 
of the stress-energy tensor $T_{\mu}^\nu$, 
$R$ is the Ricci scalar and
$\Box\equiv g^{\mu\nu}\nabla_{\mu}\nabla_{\nu}$ is the
covariant d'Alembertian, ${\nabla}_{\mu}$ being the covariant derivative
operator associated with the space-time metric $g_{\mu \nu}$. 
Furthermore,
$W=C^{\xi\sigma\mu\nu}C_{\xi\sigma\mu\nu}$ is the ``square'' of the Weyl tensor $C_{\xi\sigma\mu\nu}$ and $\mathcal{G}$  
the Gauss-Bonnet topological invariant, given by  
\begin{equation}
W=R^{\xi\sigma\mu\nu}R_{\xi\sigma\mu\nu}-2R^{\mu\nu}R_{\mu\nu}+\frac{1}{3}R^2\,,\quad 
\mathcal{G}=R^{\xi\sigma\mu\nu}R_{\xi\sigma\mu\nu}-4R^{\mu\nu}R_{\mu\nu}+R^2\,,
\end{equation}
with $R_{\mu\nu}$ the Ricci tensor and $R_{\xi\sigma\mu\nu}$ the Riemann one.
The dimensionfull coefficients $\alpha$, $\beta$, and $\xi$ of 
the above expression are related to the number of conformal fields 
present in the theory. 
We introduce real scalar fields 
$N_\mathrm{S}$, the Dirac (fermion) fields $N_\mathrm{F}$, vector fields 
$N_\mathrm{V}$, gravitons $N_2(=0\,,1)$, and higher-derivative conformal 
scalars $N_{\text{HD}}$. 
We represent $\alpha$ and $\beta$ as 
\begin{equation}
\alpha=\frac{ N_\mathrm{S}+6N_\mathrm{F}+12N_\mathrm{V}+611 N_2-8N_{\text{HD}} }{120(4\pi)^2}\,,\quad
\beta=\frac{N_\mathrm{S}+11N_\mathrm{F}+62N_\mathrm{V}+1411 N_2-28 N_{\text{HD}}}{360(4\pi)^2}\,,
\end{equation}
so that these can be positive for usual matter, 
except the higher-derivative conformal scalars. 
If we exclude the contribution of gravitons and higher-derivative conformal scalars, we get
\begin{equation}
\alpha=\frac{1}{120(4\pi)^2}(N_\mathrm{S}+6N_\mathrm{F}+12N_\mathrm{V})\,,\quad
\beta=\frac{1}{360(4\pi)^2}(N_\mathrm{S}+11N_\mathrm{F}+62N_\mathrm{V})\,,\quad
\xi=-\frac{N_\mathrm{V}}{6(4\pi)^2}\,,
\end{equation}
where we have also added the value of $\xi$~\cite{Birel, CFT2}.
In this work, it can be recognized that in terms of all the results, 
the results will be independent of the choice of 
general notation of these coefficients, 
except the fact that $\alpha$ and $\beta$ are defined to be positive. 
However, to better analyze the physical quantities, 
it is useful to have some reference value that here we give. 
For $\mathcal{N}_{\mathrm{super}} = 4$ SU(N) super Yang-Mills (SYM) theory, 
we have 
$N_\mathrm{S} = 6N^2$, $N_\mathrm{F} = 2N^2$, and $N_\mathrm{V} = N^2$, 
where $N$ is a very large number. Therefore, we obtain a relation among the numbers of scalars, spinors and vector fields. 
As a consequence, we find 
\begin{equation}
\alpha=\beta=\frac{N^2}{64\pi^2}\,,\quad\xi=-\frac{N^2}{96\pi^2}\,.
\label{setting}
\end{equation}
Note that 
\begin{equation}
\frac{2}{3}\alpha+\xi =0\,,
\label{setting2}
\end{equation}
and in principle the contribution of the $\Box R$ term to the conformal anomaly vanishes, but it could be reintroduced via a higher curvature term in the action (see below). Owing to the conformal anomaly, the 
classical Einstein equation is corrected as 
\begin{equation}
R_{\mu\nu}-\frac{1}{2}g_{\mu\nu}\,R =\kappa^2\langle T_{\mu\nu}\rangle\,. 
\label{field1}
\end{equation}
Here, we mention that $\langle T_{ij}\rangle$ is the vacuum expectation value 
of the quantum stress-energy tensor, whose trace reduces to Eq.~(\ref{TA}). 
By taking the trace of the last equation (\ref{field1}), 
we derive
\begin{equation}
R=-\kappa^2\langle T_{\mu}^\mu\rangle\equiv-\kappa^2
\left[\alpha\left(W+\frac{2}{3}\Box R\right)-\beta \mathcal{G}
+\xi\Box R\right]\,.
\label{traccia}
\end{equation}
Despite the fact that in Eq.~(\ref{setting2}), the coefficient of the $\Box R$ term is equal to zero,  
we can set it to any desired value by adding the finite $R^2$ counter term in the action. In the classical Einstein gravity, this additional term is necessary to exit from inflation~\cite{Staro}.  
Concretely, by adding the following action~\cite{Haw}
\begin{equation}
I=\frac{\gamma N^2}{192\pi^2}\int_\mathcal{M} d^4 x \sqrt{-g}\,R^2\,, \quad\gamma>0\,,
\label{R2}
\end{equation}
where $g$ is the determinant of the metric tensor,
$\mathcal M$ the space-time manifold, and $\gamma$ a positive number, 
we see that Eq.~(\ref{field1}) becomes 
\begin{equation}
R_{\mu\nu}-\frac{1}{2}g_{\mu\nu}\,R= 
-\frac{\gamma N^2\kappa^2}{48\pi^2}R R_{\mu\nu}
+\frac{\gamma N^2\kappa^2}{192\pi^2}R^2 g_{\mu\nu}
+\frac{\gamma N^2\kappa^2}{48\pi^2}\nabla_\mu\nabla_\nu R
-\frac{\gamma N^2\kappa^2}{48\pi^2}g_{\mu\nu}\Box R^2
+\kappa^2\langle T_{\mu\nu}\rangle\,. 
\end{equation}
{}From this relation, we acquire the trace 
\begin{equation}
R=
\alpha\left(W+\frac{2}{3}\Box R\right)-\beta \mathcal{G}
+\xi\Box R +\frac{\gamma N^2\kappa^2}{16\pi^2}\Box R\,.
\end{equation}
Namely, the quantity $(2/3)\alpha+\xi$ can be shifted as
\begin{equation}
\frac{2}{3}\alpha+\xi 
\rightarrow-\frac{\gamma N^2}{16\pi^2}\,.
\end{equation}
It means that when 
the Yang-Mills theory is coupled to gravity, the presence of the additional action in Eq.~(\ref{R2}) can be regarded as a higher curvature correction to the Einstein gravity or a part of the matter action.

\section{Trace anomaly in $F(R)$ gravity}

In this section, 
we investigate the trace anomaly in $F(R)$ gravity supporting the inflationary scenario. 
The action is given by\footnote{In this paper, since we use the terminology of ``$F(R)$ gravity'', the form of $F(R)$ is considered to be 
$F(R) = R+2\kappa^2\tilde\gamma R^2+f(R)$.} 
\begin{equation}
I=\frac{1}{2\kappa^2}\int_\mathcal{M} d^4 x \sqrt{-g}\,\left[R+2\kappa^2\tilde\gamma R^2+f(R)+2\kappa^2\mathcal{L}_\mathrm{QC}\right]\,,
\quad
\tilde{\gamma} \equiv \frac{\gamma N^2}{192\pi^2}\,,
\label{action}
\end{equation}
where we have considered the $R^2$ term in the action with $\tilde\gamma$ as in (\ref{R2}) and we have added a correction given by a function $f(R)$ of the Ricci scalar. Moreover, $\mathcal{L}_\mathrm{QC}$ is the matter Lagrangian of quantum corrections. 
The field equations are derived as
\begin{eqnarray}
G_{\mu\nu}\equiv R_{\mu\nu}-\frac{1}{2}g_{\mu\nu}\,R &=&
\kappa^2\langle T_{\mu\nu}\rangle 
-4\tilde\gamma\kappa^2 R R_{\mu\nu}
+\tilde\gamma R^2\kappa^2 g_{\mu\nu}
+4\tilde\gamma\kappa^2\nabla_\mu\nabla_\nu R
-4\tilde\gamma\kappa^2 g_{\mu\nu}\Box R^2
\nonumber\\
&&-f_R(R)\left(R_{\mu\nu}-\frac{1}{2}Rg_{\mu\nu}\right)+\frac{1}{2}g_{\mu\nu}[f(R)-R f_R(R)]
+(\nabla_{\mu}\nabla_{\nu}-g_{\mu\nu}\Box)f_R(R)\,.
\label{fieldequation}
\end{eqnarray}
Here, $G_{\mu\nu}$ is the Einstein tensor and $f_R(R) \equiv \partial f(R)/\partial R$. In the following, the subscription ``$R$'' indicates the derivatives with respect to the Ricci scalar. 
The trace is described as 
\begin{equation}
R=-\kappa^2 
\left(\alpha W-\beta G+\delta\Box R \right)
-2f(R)+R f_R(R)+3\Box f_R(R)\,,
\label{tr}
\end{equation}
where we have imposed the condition in Eq.~(\ref{setting2}) 
and introduced $\delta$ defined as 
\begin{equation}
\delta \equiv -12\tilde\gamma = -\frac{\gamma N^2}{16\pi^2}\,,\quad\delta<0\,. 
\label{delta}
\end{equation}
Here, $\gamma (>0)$ is a positive constant and it remains a free parameter. 

The flat FLRW space-time is described by the metric
\begin{equation}
ds^{2}=-dt^{2}+a^2(t) \left(dx^2+dy^2+dz^2\right)\,,
\label{metric}
\end{equation}
where $a(t)$ is the scale factor. In this background, we obtain
\begin{equation}
R=12H^2+6\dot H\,,
\quad 
\mathcal{G}=24H^2\left(H^2+\dot H\right)\,,
\quad 
W=0\,.
\end{equation}
Here, $H=\dot{a}(t)/a(t)$ is the Hubble parameter and 
the dot denotes the time derivative. 
The energy density $\rho$ and pressure $p$ of quantum corrections are 
represented as 
\begin{equation}
\langle T_{00} \rangle=\rho\,,\quad\langle T_{ij}\rangle=p\, a(t)^2\delta_{ij}\,,\quad (i,j=1, 2, 3)\,.
\end{equation}
In the FLRW background, 
it follows {}from $(\mu,\nu)=(0,0)$ component and the trace part of 
$(\mu,\nu)=(i,j)$ of 
Eq.~(\ref{fieldequation}), in which  
the contributes of $R^2$ term are included  in the conformal anomaly, 
we obtain the equations of motion (EoM) 
\begin{eqnarray}
\frac{3}{\kappa^{2}}H^{2} \Eqn{=} 
\rho+
\frac{1}{2\kappa^{2}}
\left[ R f_R(R)-f(R)-6H^2 f_R(R)-6H\dot f_R(R)
\right] \equiv \rho_\text{\text{eff}}\,, 
\label{EoM-00}\\ 
-\frac{1}{\kappa^{2}} \left( 2\dot H+3H^{2} \right) 
\Eqn{=} 
p+\frac{1}{2\kappa^{2}} \Bigl[
-R f_R(R)+f(R)+(4\dot H+6H^2)f_R(R)+4H\dot f_R(R)+2\ddot f_R(R)
\Bigr]\equiv p_\text{\text{eff}}\,. 
\label{EOMs}
\end{eqnarray}
In these equations,
$\rho_\text{\text{eff}}$ and $p_\text{\text{eff}}$ are the effective energy density and pressure of the universe, that is, these of total energy components 
of the universe which consist of quantum matter contents and modification of gravity. 
Thus, by considering the time-covariant derivative of (\ref{fieldequation}) and by taking account of the covariant derivative in Eq.~(\ref{fieldequation}) 
and $\nabla^\mu G_{\mu 0}=0$, we derive the effective conservation law
\begin{equation}
\dot\rho_{\text{eff}}+3H \left(\rho_{\text{eff}}+p_{\text{eff}}\right)=0\,.
\label{eq:continuity}
\end{equation}
Now, we can use this equation to get the full effective energy-momentum tensor (in the specific case, we need $\rho$ and $p$ of the quantum corrections). 
Since we have 
\begin{equation}
-\rho_{\text{eff}}+3p_{\text{eff}}=-\frac{R}{\kappa^2} \equiv 
-\beta G+\delta\Box R 
+\frac{1}{\kappa^2}\left(2f(R)-R f_R(R)-3\Box f_R(R)\right)\,,
\end{equation}
where we have used Eq.~(\ref{tr}) with $W=0$. 
Eliminating the pressure and using the conservation law in Eq.~(\ref{eq:continuity}), we find
\begin{equation}
\frac{d}{dt}\left(\rho_{\text{eff}} a^4\right)=-\dot a a^3\left(-\rho_{\text{eff}}+3p_{\text{eff}}\right)
\equiv
a^3\dot a\left[24\beta\frac{\dot a^2\ddot a}{a^3}
+\delta \left(\ddot{R}+3H\dot{R}\right)\right] 
-\frac{\dot{a} a^3}{\kappa^2}\left(2f(R)-R f_R(R)-3\Box f_R(R)\right)\,.
\end{equation}
The integration of this equation yields the expression for the effective energy density 
\begin{equation}
\rho_{\text{eff}}=\frac{\rho_0}{a^4}+6\beta H^4+\delta\left(18H^2\dot H+6\ddot H H-3\dot H^2\right)
+\frac{1}{2\kappa^2}\left(R f_R(R)-f(R)-6H^2 f_R(R)-6H \dot f_R(R)\right)\,, 
\label{rho}
\end{equation}
where $\rho_0$ is the constant of integration. 
Hence, eventually we have the effective pressure
\begin{eqnarray}
p_\text{eff} \Eqn{=} 
\frac{\rho_0}{3a^4}-\beta \left(6H^4+8H^2\dot H\right)
-\delta\left(
9\dot H^2+12H\ddot H+2\dddot H+18H^2\dot H
\right)+
\nonumber\\&&
\frac{1}{2\kappa^{2}} \Bigl[
-R f_R(R)+f(R)+(4\dot H+6H^2)f_R(R)+4H\dot f_R(R)+2\ddot f_R(R)
\Bigr]\,.
\label{p}
\end{eqnarray}
In the expressions of $\rho_\text{eff}$ in Eq.~(\ref{rho}) 
and $p_\text{eff}$ in Eq.~(\ref{p}), 
we can recognize the contributions from not only modified gravity 
but also quantum corrections. 
The appearance of the integration constant $\rho_0$ implies that 
the quantum state may contain an arbitrary amount of radiation~\cite{Haw}, 
which does not contribute to the trace of the conformal anomaly. 
In the seminal paper by Starobinsky~\cite{Staro}, 
$f(R)=0$, the model is generalized to a non-necessarily flat spatial 
curvature and $\rho_0=0$ (in comparison with the coefficients $k_1$ and $k_2$ used in Ref.~\cite{Staro}, we have $k_2=90N^2$ and $k_3=-180N^2\gamma<0$). 
We also put 
\begin{equation}
\rho_0=0\,,
\label{00}
\end{equation}
motivated by the fact that we are interested in the Planck energy scale, 
where the energy density of the standard radiation 
can be much smaller than those of 
the quantum corrections and modification of gravity at this scale, 
and therefore the contribution from radiation can be neglected. 

Here, we present several remarks. 
In principle, the form of the stress energy tensor of quantum corrections is unknown. What we know is its trace introduced in the general form in Eq.~(\ref{TA}). However, given the metric (the flat FLRW space-time in our case), it is possible to derive the stress-energy tensor of the quantum anomaly ``on shell'', 
as executed above. If we know the effective energy density of the universe due to the quantum corrections and modified gravity, the pressure follows from the conservation law which is automatically satisfied. For this reason, in order to study the cosmological evolution of the model, it is enough to use one of the gravitational equations (\ref{EoM-00}) and (\ref{EOMs}). In what follows, we use the Friedmann equation (\ref{EoM-00}), where $\rho_\mathrm{eff}$ is given by Eq.~(\ref{rho}) with $\rho_0=0$.

\section{Trace-anomaly driven inflation in exponential gravity}

To reproduce the cosmic acceleration of the de Sitter universe today, 
many $f(R)$ gravity models have been presented in the literature.
The simplest idea is to construct a function of $f(R)$ which 
mimics the cosmological constant $\Lambda$ for large curvature and goes to zero for $R=0$. Such behaviour can be sketched by the Heaviside function $\theta \left(R-R_*\right)$ as
\begin{equation}
f(R)=-2\Lambda\theta\left(R-R_*\right)\,,
\label{proto1}
\end{equation}
where $R_*$ is a fixed curvature smaller than the current curvature. 
It is intuitive that in this way, we can recover the $\Lambda$CDM model at large curvature and the Minkowski space-time for $R=0$. 
In Refs.~\cite{HuSaw,Nojiri:2007as,Starobinsky:2007hu,Cognola:2007zu,Linder:2009jz,Battye}, 
several versions of this kind of (viable) $F(R)$ gravity have been 
proposed. 
For the inflation, one also needs a (quasi)-de Sitter solution and the introduction of a suitable effective cosmological constant at large curvature appears to be quite natural, such that the model above can be implemented as 
\begin{equation}
f(R)=-2\Lambda\theta\left(R-R_*\right)-2\Lambda_\text{eff}\theta\left(R-R_0\right)\,,
\label{proto2}
\end{equation}
where $\Lambda_\text{eff}$ and $R_0$ are constants and 
$\Lambda_\text{eff}\,, R_0\gg \Lambda\,, R_*$. 
However, the most significant problem of inflationary cosmology is how to realize a graceful exit from inflation. 
By making use of an exponential model in Eq.~(\ref{proto2}), where 
two effective cosmological constants are incorporated, 
an attempt to unify inflation with the late-time cosmic acceleration 
has been analyzed in detail in Ref.~\cite{twosteps}. 
In such a case, some power-law term of the Ricci scalar has been added 
to make the exit from inflation possible 
by taking into consideration the perturturbations coming from ultrarelativistic matter and radiation present in the hot universe scenario. 
On the other hand, in this work we analyze the effects of the trace anomaly in modified gravity theories, that is, we study physics at the Planck scale, so that contributions of standard matter/radiation will be negligible. 
As a toy model of form of $f(R)$, we examine 
\begin{equation}
f(R)=-2\Lambda_{\text{eff}}\left[1- \exp\left(-\frac{R}{R_0}\right)\right]\,. 
\label{model}
\end{equation}
For this model, 
in the small curvature limit, the modification of gravity vanishes, so that 
the cosmic evolution can be the same as the standard scenario, 
whereas at large curvature (i.e., the energy scale of inflation) $R\gg R_0$, 
an effective cosmological constant $\Lambda_\text{eff} (>0)$ appears. 
Beside its simplicity, 
the reason why we consider this model is as follows.  
Since the conformal anomaly leads to the de Sitter solution, 
the adding of a constant to the effective energy density of the universe only shifts the de Sitter solution of inflation without cancelling it. 
Hence, in this model, the perturbations around the leading terms of $f(R)$ proportional to $\exp\left(-R/R_0\right)$ are negative, so that for the (unstable) de Sitter solution, the Hubble parameter and therefore the Ricci scalar slowly decrease during inflation, as is seen later. {}From now on, we analyze the cosmological behaviour in this model with the quantum anomaly.

\subsection{De Sitter solution(s)}

Since inflation is described by the de Sitter solution, first we 
investigate the de Sitter solutions of the model in Eq.~(\ref{model}).
At large curvature $R\gg R_0$, the model (\ref{model}) reduces to
$f(R)\simeq-2\Lambda_\text{eff}$, and the Friedmann equation (\ref{EoM-00}) 
reads
\begin{equation}
\frac{3}{\kappa^2}H^2\simeq 6\beta H^4+\delta\left(18 H^2\dot H+6\ddot H H -3\dot H^2\right)+\frac{\Lambda_\text{eff}}{\kappa^2}\,,\label{F1}
\end{equation}
where we have omitted the terms proportional to $\exp\left(-R/R_0\right)$. 
This equation (with $\Lambda_\text{eff}\neq 0$) has two de Sitter solutions 
with the constant Hubble parameter
\begin{equation}
H_{\text{dS}\pm}^2=\frac{1}{4\beta\kappa^2}\left[1\pm\sqrt{1-\frac{8\Lambda_\text{eff}\beta\kappa^2}{3}}\right]\,,
\label{+-}
\end{equation}
which are independent of the parameter $\delta$ 
(it is well know that the $R^2$ term does not give any contribution to the de Sitter solution). Since $\beta =N^2/\left( 64\pi^2 \right) >0$ 
in Eq.~(\ref{setting}), 
in order to have real (and positive) solutions, the following condition has 
to be required: 
\begin{equation}
\Lambda_\text{eff}<\frac{3}{8\beta\kappa^2} \,, 
\quad 
\frac{1}{\beta\kappa^2} \equiv 
\frac{8\pi M_{\text{Pl}}^2}{N^2}\,,
\label{condlamb}
\end{equation}
where we have reintroduced the Planck mass. 
Solutions in Eq.~(\ref{+-}) can be rewritten as
\begin{equation}
H_{\text{dS}\pm}^2=\frac{1}{4\beta\kappa^2}\left(1\pm\sqrt{1-\frac{8\zeta}{3}}\right) = 
\frac{2\pi M_{\text{Pl}}^2}{N^2}
\left(1\pm\sqrt{1-\frac{8\zeta}{3}}\right)\,, 
\quad 
\Lambda_\text{eff}=\frac{\zeta}{\beta\kappa^2} =
\zeta\left[\frac{8\pi M_{\text{Pl}}^2}{N^2}\right]\,, 
\quad 0<\zeta<\frac{3}{8}\,.
\label{Hpm}
\end{equation}
Furthermore, there are two special solutions
\begin{eqnarray}
H_\text{dS}^2 \Eqn{=} \frac{1}{2\beta\kappa^2} = \frac{4\pi M_{\text{Pl}}^2}{N^2}\,,\quad \Lambda_\text{eff}=0\,, 
\label{Hun-00} \\
\quad H_\text{dS}^2 \Eqn{=} \frac{1}{4\beta\kappa^2} = \frac{2\pi M_{\text{Pl}}^2}{N^2}\,,\quad\Lambda_\text{eff}=\frac{3}{8\beta\kappa^2} = \frac{3}{8}\left(\frac{8\pi M_{\text{Pl}}^2}{N^2}\right)\,.
\label{Hun}
\end{eqnarray}
The first solution in Eq.~(\ref{Hun-00}) 
corresponds to the one obtained in the Starobinsky model 
in Ref.~\cite{Staro}. 
We here remark several points on the condition in Eq.~(\ref{condlamb}). 
Even if $N\gg 1$, the effective cosmological constant of the model (\ref{model}) is bounded 
at the very large value of the Planck length 
($l_\text{Pl}=M_\text{Pl}^{-1}$). The corresponding energy density is given by 
$\rho_{\Lambda_\text{eff}}=\Lambda_\text{eff}/\kappa^2\propto M^4_\text{Pl}$. 
The fact that the energy density associated with the de Sitter solution is of 
order of the Planck scale to the fourth power justifies the introduction of 
the quantum corrections into the theory.
Moreover, in the present derivation we have assumed that both the de Sitter solutions $H_{\text{dS}\pm}^2$ are acquired in the high curvature limit of the model in Eq.~(\ref{model}), namely, at $R_{\text{dS}\pm} (\equiv 12H_{\text{dS}\pm}^2)\gg R_0$. 

We study the stability of the de Sitter solutions found above. 
We define the perturbations $\Delta H(t)$ as 
\begin{equation}
H=H_{\text{dS}\pm}+\Delta H(t)\,, 
\quad |\Delta H(t)|\ll 1\,.
\end{equation}
Substituting this expression into Eq.~(\ref{F1}) and 
taking the first order of $\Delta H(t)$, we find 
\begin{equation}
\Delta \ddot{H}(t)+3H_{\text{dS}\pm}\Delta \dot{H}(t) 
=\frac{\Delta H(t)}{\delta}\left(\frac{1}{\kappa^2}-4H_{\text{dS}\pm}^2\beta\right)\,.
\label{eq:p-eq}
\end{equation}
Here, we have not included any term proportional to $\exp \left(-R/R_0\right)$ 
of the model in Eq.~(\ref{model}), which contributes to the stability of the solution only through $\Lambda_\text{eff}$ inside the expression of 
$H_{\text{dS}\pm}$ in Eq.~(\ref{+-}). 
The solution of Eq.~(\ref{eq:p-eq}) is given by 
\begin{equation}
\Delta H(t)=A_0\text{e}^{\lambda_{1,2} t}\,,\quad
\lambda_{1,2}=\frac{-3H_{\text{dS}\pm}\pm\sqrt{9H_{\text{dS}\pm}^2+\frac{4}{\delta}\left(\frac{1}{\kappa^2}-4H_{\text{dS}\pm}^2\beta\right)}}{2}\,,
\label{pert}
\end{equation}
where $A_0$ is a constant. 
The de Sitter solutions of the model (\ref{model}) are unstable (and adopted to describe the inflation) 
only if 
$\lambda_1$ (the eigenvalue with the positive sign in front of the square root) is real and positive, i.e., 
\begin{equation} 
4\beta-\frac{1}{\kappa^2H_{\text{dS}\pm}^2} >0\,,
\quad 
9H_{\text{dS}\pm}^2+\frac{4}{\delta}\left(\frac{1}{\kappa^2}-4H_{\text{dS}\pm}^2\beta\right)>0\,.
\label{expinst}
\end{equation}
Here, we have taken into account the fact that $\beta>0$ and $\delta<0$. 
We further analyze the results.

\subsubsection{Double de Sitter solution}

By plugging $H_{\text{dS}\pm}$ in Eq.~(\ref{Hpm}) into Eq.~(\ref{expinst}), 
we see that $\lambda_1$ is real if
\begin{equation}
1\pm\left(1-\frac{16\beta}{9\delta}\right)\sqrt{1-\frac{8}{3}\zeta}>0\,,
\label{im}
\end{equation}
where the plus and minus signs correspond to the ones of $H_{\text{dS}\pm}$. 
In the case of $H_{\text{dS}+}$, this condition is always satisfied, and 
eventually the instability condition reads
\begin{equation}
\sqrt{1-\frac{8}{3}\zeta}>0\,. 
\end{equation}
Hence, this de Sitter solution is unstable effectively. 
On the other hand, 
in the case of $H_{\text{dS}-}^2$, the condition in Eq.~(\ref{im}) is satisfied only if 
\begin{equation}
\frac{1}{1-\frac{16\beta}{9\delta}}>\sqrt{1-\frac{8}{3}\zeta}\,.
\end{equation}
However, even if this condition could be satisfied, from Eq.~(\ref{expinst}) 
we have 
\begin{equation}
-\sqrt{1-\frac{8}{3}\zeta}<0\,,
\end{equation}
and thus the de Sitter solution becomes stable. 
Consequently, 
a stable de Sitter solution can be realized by the model (\ref{model}) during the inflation (and it may be the final attractor of the system), 
and therefore inflation cannot end. 
Next, we investigate 
the second de Sitter solution in the small curvature limit of the model in 
Eq.~(\ref{model}) 
\begin{equation}
R_{\text{dS}-}\leq R_0\ll R_{\text{dS}+}\,,
\label{Rcond}
\end{equation}
where $R_{\text{dS}\pm}=12 H_{\text{dS}\pm}^2$. 
In this case, the second (stable) de Sitter solution cannot be obtained 
with the mechanism illustrated above. 
In what follows, we identify the de Sitter solution with $H_{\text{dS}}\equiv H_{\text{dS}+}$, considering the relation (\ref{Rcond}) satisfied. This de Sitter solution is unstable and describes inflation in the model (\ref{model}).
We note that since the Starobinsky model is free of singularities 
in both the past and the future, also our theory results in being free of singularities, where an effective cosmological constant is added to the conformal anomaly.

\subsubsection{Special de Sitter solutions}

For the special solutions in Eqs.~(\ref{Hun-00})--(\ref{Hun}), we see that in the first case (the Starobinsky model), from Eq.~(\ref{expinst}) we have 
\begin{equation}
2\beta>0\,,
\quad
\frac{9}{2\beta}-\frac{4}{\delta}>0\,,
\end{equation}
and it is well-known that the solution is unstable. 
For the second solution, we get
\begin{equation}
4\beta-\frac{1}{\kappa^2H_{\text{dS}}^2}=0\,,
\quad
\frac{9}{4\beta\kappa^2}>0\,,
\end{equation}
and a later analysis is required. In what follows, we avoid to discuss these two special cases, namely, we assume $\Lambda_\text{eff}\neq 0$ and  $\Lambda_\text{eff}\neq 3/(8\beta\kappa^2)$ in the model (\ref{model}). 

Here, we mention that 
in the case of exponential gravity, 
the exponential term makes 
the possibility for the de Sitter solution to be 
unstable wider than it for the pure trace-anomaly driven inflation. 
As a result, we may build the model in Eq.~(\ref{model}) in order to meet all 
the recent observational data in terms of inflation. This cannot be realized 
by pure trace-anomaly only (see Sec.~IV B). In other words, a modification of 
gravity from general relativity in the early universe may be necessary 
to have a viable dynamics of inflation. 

\subsection{Dynamics of inflation}

Given the unstable de Sitter solution $H_{\text{dS}\pm}^2$ in (\ref{+-}), 
to analyze inflation occurring in the model in Eq.~(\ref{model}), 
we have to calculate the amplitude of the perturbations in Eq.~(\ref{pert}), that is, we need the bounding value of $A_0$. In particular, its sign is crucial because the evolution of the Hubble parameter and hence the cosmological behaviour of the universe depend on it. 
At the time $t=0$ when inflation starts, we have to set 
$\Delta H(t=0)= 0$. {}From the Friedmann equation~(\ref{EoM-00}), 
by taking into account also the terms in the model (\ref{model}) which are of the order of 
$\exp\left(-R_\text{dS}/R_0\right)$, around the de Sitter solution 
$H=H_\text{dS}+\Delta H(t)$, we find 
\begin{equation}
-\delta\left(\Delta\ddot H(t)+3H_{\text{dS}}\Delta \dot H(t)\right)+
\Delta H(t)\left(\frac{1}{\kappa^2}-4H_{\text{dS}}^2\beta\right)=-\frac{\text{e}^{-R_\text{dS}/R_0}\Lambda_\text{eff}}{12H_\text{dS}\kappa^2}
\left(
\frac{R_\text{dS}}{R_0}+2
\right)\,.\label{ciccio}
\end{equation}
The complete solution of this equation is given by the homogeneous part in Eq.~(\ref{pert}) plus the contribute of modified gravity as follows 
\begin{equation}
\Delta H(t)=A_0\text{e}^{\lambda_{1,2} t}
-\frac{\text{e}^{-R_\text{dS}/R_0}\Lambda_\text{eff}}{12H_\text{dS}\kappa^2}
\left(
\frac{R_\text{dS}}{R_0}+2
\right)\left(\frac{1}{\kappa^2}-4H_{\text{dS}}^2\beta\right)^{-1}\,.
\label{***}
\end{equation}
Thus, at $t=0$, by putting $\Delta H(t=0)= 0$, we can estimate the amplitude $A_0$ as 
\begin{equation}
A_0=\frac{\text{e}^{-R_\text{dS}/R_0}\Lambda_\text{eff}}{12H_\text{dS}\kappa^2}
\left(
\frac{R_\text{dS}}{R_0}+2
\right)\left(\frac{1}{\kappa^2}-4H_{\text{dS}}^2\beta\right)^{-1}\,,
\end{equation}
which leads to
\begin{equation}
A_0=-\frac{\text{e}^{-R_\text{dS}/R_0}\zeta}{12H_\text{dS}(\beta\kappa^2)}
\left(
\frac{R_\text{dS}}{R_0}+2
\right)\left(1-\frac{8}{3}\zeta\right)^{-1/2}<0\,.\label{A0}
\end{equation}
Here, we have considered only the unstable solution $H_\text{dS}\equiv H_\text{dS+}$ in Eq.~(\ref{Hpm}) as discussed above. 
Note that thanks to Eq.~(\ref{A0}), all the terms omitted in Eq.~(\ref{ciccio}) are at lower orders, particularly, $|(\Delta H(t)/H_\text{dS})^2|\ll \exp\left(-R_\text{dS}/R_0\right)$. Moreover, as the time passes, the second term in Eq.~(\ref{***}) can be neglected in comparison with the first one and eventually we recover Eq.~(\ref{pert}). 
This point is quite important in our analysis. 
In principle, in an $F(R)$ gravity theory, it is possible to judge 
whether a given solution is stable or unstable. 
In the de Sitter solution, we acquire a contribution originating from the modification of gravity, but the instability of the solution only depends on the trace anomaly (in fact, it has the dependence on the positive $R^2$ correction term to the Einstein gravity due to the term $\delta (< 0)$). 
However, it is crucial issue to understand the origin of the perturbations during this phase to bound them at the beginning of inflation. 
If a modified gravity theory describes the early universe and relates to the quantum anomaly, it may generate the perturbations to the (unstable) de Sitter solution (otherwise, if $f(R)=-2\Lambda_\text{eff}$, the de Sitter solution is an exact solution of the gravitational field equations), which grow up in time and 
realize a graceful exit from inflation. 

In the case of model (\ref{model}), the sign of amplitude of the perturbations is negative. The standard cosmology is recovered at small curvature ($R/R_0\ll 1$), where it is expected that the modifications to the Einstein gravity vanish and the expansion of the universe is driven by matter/radiation. Inflation ends when the perturbation $H_\text{dS}|\Delta H(t)|$ is of the order of $H_\text{dS}^2$, i.e., 
\begin{equation}
\exp\left(\lambda_1 t-\frac{R_\text{dS}}{R_0}\right)\simeq\frac{3}{\zeta}\left(1-\frac{8}{3}\zeta\right)\left(1+\sqrt{1-\frac{8}{3}\zeta}\right)\left(\frac{R_\text{dS}}{R_0}+2\right)\,.
\end{equation}
As a consequence, we may estimate the time at the end of inflation as
\begin{equation}
t_\mathrm{f} \simeq\frac{R_\text{dS}}{R_0\,\lambda_1}\,.\label{tf}
\end{equation}
Note that since $\lambda_1\sim H_\text{dS}$, this time is at the Planck scale. 

The primordial acceleration can solve the problem of initial conditions of the standard model (horizon and velocities problems), only if $\dot a_\mathrm{i}/\dot a_0< 10^{-5}$, where $\dot a_\mathrm{i}\,,\dot a_0$ are the time derivatives of the scale factor at the Big Bang and today, respectively, and $10^{-5}$ is the estimated value of the inhomogeneous perturbations in our universe. Since in decelerating universe $\dot a(t)$ only decreases of a factor $10^{28}$, it is required that $\dot a_\mathrm{i}/\dot a_\mathrm{f}<10^{-33}$, where $a_\mathrm{i}$ is the scale factor at the Big Bang again (namely, at the beginning of inflation), and $a_\mathrm{f}$ is the scale factor at the end of inflation. Thus, if inflation is governed by a (quasi) de Sitter solution where $a(t)=\exp\left(H_\text{dS}t\right)$, we can introduce the number of $e$-folds $\mathcal N$ as
\begin{equation}
\mathcal N=\ln \left(\frac{a_\mathrm{f}}{a_\mathrm{i}}\right)\,,
\end{equation}
and inflation is viable if $\mathcal{N}>76$. 
For the model (\ref{model}), by taking account of the fact that we have chosen 
$t_\mathrm{i}=0$ and using Eq.~(\ref{tf}), we acquire 
\begin{equation}
\mathcal N\equiv H_\text{dS} t_f=\frac{2R_\text{dS}}{3 R_0}\left[-1+\sqrt{1-\frac{16\beta}{9\delta}\left(\frac{\sqrt{1-\frac{8}{3}\zeta}}{1+\sqrt{1-\frac{8}{3}\zeta}}\right)}\right]^{-1}\,.
\label{N}
\end{equation}
For simplicity, we represent
\begin{equation}
R_\text{dS}=b R_0\,,
\quad 
b>0\,,
\label{b}
\end{equation}
where $b$ is a positive number. 
By combining this relation, 
the expressions for $\beta$ in Eq.~(\ref{setting}) and 
$\delta$ in Eq.~(\ref{delta}), and Eq.~(\ref{N}), we have 
\begin{equation}
\mathcal N=\frac{2 b}{3}\left[-1+\sqrt{1+\frac{4}{9\gamma}\left(\frac{\sqrt{1-\frac{8}{3}\zeta}}{1+\sqrt{1-\frac{8}{3}\zeta}}\right)}\right]^{-1}\,.\label{Nf}
\end{equation}
We remember that $\gamma$ enters the action of the theory in front of the $R^2$ term in Eq.~(\ref{action}) and $\zeta$ depends on the effective cosmological constant of the model under investigation as in Eq.~(\ref{Hpm}). To satisfy Eq.~(\ref{Rcond}) and not to have the stable de Sitter solution of the model (\ref{model}), the following relation has to be met
\begin{equation}
1\ll b\leq\frac{1+\sqrt{1-\frac{8}{3}\zeta}}{1-\sqrt{1-\frac{8}{3}\zeta}}\,.
\label{eq:COND-00}
\end{equation}
As an example, for $\Lambda_\text{eff}=\pi M_\text{Pl}^2/N^2$, that is, 
$\zeta=1/8$ and $b=3$, the condition in Eq.~(\ref{eq:COND-00}) is satisfied. 
Moreover, 
since $R_{\text{dS}-}<R_0 (\simeq 3R_{\text{dS}-})$, it cannot be met ($f(R_{\text{dS}-})\neq-2\Lambda_\text{eff}$), 
whereas since there are the relations $R_{\text{dS}+}=3R_0$ and $\exp\left(-R_{\text{dS}+}/R_0\right) \sim 0.05 \ll 1$, we find $f(R_{\text{dS}+})\simeq -2\Lambda_\text{eff}$, and therefore the unstable de Sitter solution $R_\text{dS}\equiv R_{\text{dS}+}$ can be realized. 
In this case, to obtain $\mathcal N>76$, $\gamma$ has to meet 
the relation $\gamma>3.8$. 
This is an example for the setting of the parameters in the model in 
Eq.~(\ref{model}) in order to reproduce the early-time cosmic acceleration 
with an appropriate duration of inflation. However, other conditions must be 
satisfied to have a viable inflation, as is seen below. 

\subsubsection{EoS parameter}

The effective EoS parameter is defined as 
\begin{equation}
w_\text{eff} \equiv \frac{p_\text{eff}}{\rho_\text{eff}}\,,
\end{equation}
where $p_\text{eff}$ and $\rho_\text{eff}$ are 
given by Eqs.~(\ref{rho}) and (\ref{p}), respectively. 
For the de Sitter solution, $w_\text{eff} = -1$, 
it is equal to minus one, 
but by introducing the perturbation $H=H_\text{dS}+\Delta H(t)$, we obtain 
\begin{equation}
w_\text{eff}\simeq-1-\frac{2\kappa^2}{3H_\text{dS}^2}\left(4\beta H_\text{dS}^2\Delta\dot H(t)
+3\delta H_\text{dS}\Delta\ddot H(t)+\delta \Delta\dddot H(t)\right)\,.
\end{equation}
By describing $\Delta(t)=A_0\text{e}^{\lambda_1 t}$, where $\lambda_1$ is given by Eq.~(\ref{pert}) in the case of model (\ref{model}), we get
\begin{equation}
w_\text{eff}\simeq-1
-\frac{2\lambda_1 A_0\text{e}^{\lambda_1 t}}{3H_\text{dS}^2}
=-1-\frac{2\lambda_1 \Delta(t)}{3H_\text{dS}^2}\,.
\end{equation}
Thus, by using $H_\text{dS}\equiv H_{\text{dS}+}$ in Eq.~(\ref{Hpm}) and introducing the number of $e$-folds $\mathcal N$ in Eq.~(\ref{N}), 
we find the effective EoS parameter for the model (\ref{model}) during inflation, namely 
\begin{equation}
w_\text{eff}\simeq-1-\frac{2 R_\text{dS}}{3 R_0\mathcal N} 
\frac{\Delta H(t)}{H_\text{dS}}\,.
\end{equation}
The effective EoS parameter tends to increase 
because $A_0 < 0$ as shown in Eq.~(\ref{A0}). 
In addition, since $|\Delta H(0)/H_\text{dS}(0)|\ll 1$ and 
$\Delta H(t_\mathrm{f})\simeq -H_\text{dS}(t_\mathrm{f})$, 
we see that during inflation, the following relation is met 
\begin{equation}
-1<w_\text{eff}<-1+\frac{2 R_\text{dS}}{3 R_0\mathcal N}\,. 
\end{equation}
In general, $\mathcal N>R_\text{dS}/R_0 \, (=b)$, and the expansion of the universe is accelerating until the end of this period driven by a fluid with 
negative pressure. As the energy density of the fluid  
slowly decreases, and finally the exit from inflation can occur 
as seen above.

\subsubsection{Spectral index}

The second time derivative of $a(t)$ is given by 
\begin{equation}
\frac{\ddot{a}}{a}= H^2+\dot{H}=H^2\left(1-\epsilon \right)\,,
\end{equation}
where we have introduced the parameter $\epsilon$. 
When the approximate de Sitter solution is realized, 
it has to be very small as 
\begin{equation}
\epsilon=-\frac{\dot{H}}{H^2}\ll 1\,.
\end{equation}
Moreover, $\epsilon$ has to change very slowly. 
There is another parameter $\eta$, which has to also be very small as 
\begin{equation}
\left|\eta\right|=\left|-\frac{\ddot{H}}{2H\dot{H}}\right|\equiv\left|\epsilon-\frac{1}{2 \epsilon H}\dot\epsilon\right|\ll 1 \,.
\end{equation}
These two parameters are the so-called slow-roll parameters. 
In standard inflation models driven by a scalar field, called the inflaton, 
these are useful to study the viability of inflation models. 

The amplitude of scalar-mode power spectrum of 
the primordial curvature perturbations 
at $k=0.002 \, \mathrm{Mpc}^{-1}$ 
is described as 
\begin{equation}
\Delta_{\mathcal R}^2=\frac{\kappa^2 H^2}{8\pi^2\epsilon}\,,
\label{spectrum}
\end{equation}
and the last cosmological data constrain the spectral index $n_\mathrm{s}$ and the tensor-to-scalar ratio $r$ are given by~\cite{Mukhanov:1981xt, L-L} 
\begin{equation}
n_\mathrm{s}=1-6\epsilon+2\eta\,,
\quad 
r=16\epsilon\,.
\label{indexes}
\end{equation}
In the model (\ref{model}), we find 
\begin{equation}
\Delta_{\mathcal R}^2=\frac{1}{32\pi^2\beta\epsilon}\left(1+\sqrt{1-\frac{8}{3}\zeta}\right) 
=\frac{2}{N^2\epsilon}\left(1+\sqrt{1-\frac{8}{3}\zeta}\right)\,,
\end{equation}
where in the second equality, we have used Eq.~(\ref{setting}). 
The parameters $\epsilon$ and $\eta$ read 
\begin{eqnarray}
\epsilon \Eqn{\simeq} -\frac{\Delta \dot H(t)}{H_\text{dS}^2}=
\frac{b^2}{\mathcal N^2}\left(-\frac{\delta}{4\beta}\right)\frac{\text{e}^{(\lambda_1 t-b)}\zeta\left(b+2\right)}{\left(1-\frac{8}{3}\zeta\right)}\left(\frac{b}{3\mathcal N}+1\right) 
= \frac{b^2}{\mathcal N^2}\frac{\text{e}^{(\lambda_1 t-b)}\zeta\left(b+2\right)}{\left(1-\frac{8}{3}\zeta\right)}\left(\frac{b}{3\mathcal N}+1\right)\,,
\label{eq:EPS}
\nonumber\\
\eta \Eqn{=} \epsilon-\frac{\dot\epsilon}{2\epsilon H_\text{dS}}=\epsilon-\frac{\lambda_1}{2 H_\text{dS}}
= \epsilon-\frac{b}{2\mathcal N}\,. 
\label{eq:ETA}
\end{eqnarray} 
In deriving the second and third equalities in Eq.~(\ref{eq:EPS}), 
we have used the number of $e$-folds in Eq.~(\ref{Nf}), 
and in obtaining the last equality in Eq.~(\ref{eq:ETA}), 
we have taken into consideration $\mathcal N=H_\text{dS} t_\mathrm{f}$, 
where $t_\mathrm{f}$ is given by Eq.~(\ref{tf}) with $R_\text{dS}=b R_0$.
During inflation, when $t\ll t_\mathrm{f}$, since $\mathcal N\gg 1$, 
we have
\begin{equation}
\epsilon\simeq
\frac{b^2}{\mathcal N^2}\frac{\text{e}^{-b}\zeta\left(b+2\right)}{\left(1-\frac{8}{3}\zeta\right)}\ll 1\,,
\quad 
\left|\eta\right|\simeq \left|-\frac{b}{2\mathcal N}\right|\ll 1\,.
\end{equation}
Thus, the spectral index and the tensor-to-scalar ratio in Eq.~(\ref{indexes}) for the model (\ref{model}) are derived as 
\begin{equation}
n_\mathrm{s}=1-\frac{b}{\mathcal N}-\frac{6 b^2}{\mathcal N^2}\frac{\text{e}^{-b}\zeta\left(b+2\right)}{\left(1-\frac{8}{3}\zeta\right)}\,,\quad
r=\frac{16 b^2}{\mathcal N^2}\frac{\text{e}^{-b}\zeta\left(b+2\right)}{\left(1-\frac{8}{3}\zeta\right)}\,.
\end{equation}

We mention the recent observations of the spectral index $n_\mathrm{s}$ 
as well as the tensor-to-scalar ratio $r$.  
The results observed by the Planck satellite are 
$n_{\mathrm{s}} = 0.9603 \pm 0.0073\, (68\%\,\mathrm{CL})$ and 
$r < 0.11\, (95\%\,\mathrm{CL})$~\cite{Ade:2013lta}. 
Since $b/\mathcal N\ll 1$ and $1\ll b$, the constraints from the Planck satellite described above can be satisfied. 
For instance, for $b=3$, $\zeta=1/8$, and $\mathcal N=76$, we have $n_\mathrm{s} \simeq 0.9601$ and $r=1.20 \times 10^{-3}$. 

On the other hand, very recently, 
the BICEP2 experiment has detected the $B$-mode polarization of the cosmic microwave background (CMB) radiation with the tensor to scalar ratio 
$r =0.20_{-0.05}^{+0.07}\, (68\%\,\mathrm{CL})$~\cite{Ade:2014xna},  
and also the case that $r$ vanishes 
has been rejected at $7.0 \sigma$ level. 
Thus, the viable models of inflation have to produce such a finite value of 
$r$. Several attempts to build inflationary models or other cosmological 
processes to realize the value of $r=0.20$ have recently been executed, e.g., 
in Refs.~\cite{Bicep2-inflation, T-D-B, DM-CP}. 

Here, we examine the case of our model to explain the BICEP2 result. 
For our model, even if the dependence of the tensor-to-scalar ratio on $\mathcal N^2$ makes it very small, we can play with a value of $\zeta$ close to $3/8$ in order to increase its value. 
For instance, with the choice $\zeta=0.37125$, we can still describe the unstable de Sitter solution for $b>1$, since $R_\text{dS}\gg R_0$ and $f(R_\text{dS})\simeq -2\Lambda_\text{eff}$. Thus, the number of $e$-folds $\mathcal N$ depends on $\gamma$ only as in Eq.~(\ref{Nf}). 
Indeed, when we take the combination of the values of $b$ and $\gamma$, e.g., 
$(b=2, \gamma>1.14)$, $(b=3, \gamma>0.76)$, and $(b=4, \gamma>0.57)$, 
and so on, we obtain $\mathcal N>76$. 
In such cases, the value of the tensor-to-scalar ratio results in being 
non-null as implied by the BICEP2 data. 
For example, if $\mathcal N=76$,  
for $b= 2,\, 3$ and $4$, 
we acquire $r=0.22,\, 0.23$, and $0.18$, respectively. 
As a result, it is clearly seen that in our model, the BICEP2 result 
in terms of the tensor-to-scalar ratio $r$ can be realized.

\subsection{End of inflation and the following reheating stage}

At the end of inflation, when $R/R_0\ll 1$ in the model (\ref{model}), the effective energy density in Eq.~(\ref{rho}) reads at the second order of $\Delta H(t)$ as 
\begin{equation}
\rho_\text{eff}=\delta\left(6\Delta\ddot H(t) \Delta H(t)-3\Delta \dot H(t)^2\right)
+\frac{6\Lambda_\text{eff}}{\kappa^2}
\left(
\frac{3\Delta\dot H(t)^2}{R_0^2}+\frac{\Delta H(t)^2}{R_0}-\frac{6\Delta H(t)\Delta\ddot H(t)}{R_0^2}
\right)\,,
\label{Rll}
\end{equation}
where we have expanded the exponential model around $R\simeq\Delta R(t)$, with $|\Delta R(t)|\ll 1$, and 
we have considered 
$\Delta R(t)=\left(12\Delta H(t)^2+6\Delta\dot H(t)\right)$. 
Hence, from the Friedmann equation (\ref{EoM-00}), we derive 
\begin{equation}
\Delta H(t)^2\left(\frac{1}{\kappa^2}-\frac{2\Lambda_\text{eff}}{R_0\kappa^2}
\right)=
\Delta\ddot H(t)\Delta H(t)\left(2\delta-\frac{12\Lambda_\text{eff}}{\kappa^2 R_0^2}\right)
+\Delta\dot H(t)^2\left(\frac{6\Lambda_\text{eff}}{R_0^2\kappa^2}
-\delta\right)\,. 
\end{equation}
Its solution is given by
\begin{equation}
\Delta H(t)=c_0\cos\left(B_0 t
\right)^2\,,
\quad
B_0=\frac{1}{2}\sqrt{\frac{R_0(R_0-2\Lambda_\text{eff})}{6\Lambda_\text{eff}-\delta R_0^2\kappa^2}}\,,
\label{Delta0}
\end{equation}
with $c_0$ a constant. 
When we examine the evolution of the system which tends to $R=+0$ for the model (\ref{model}), 
the following relation is required: 
\begin{equation}
R_0>2\Lambda_\text{eff}\,,
\label{RL}
\end{equation}
where we have taken into account that $\delta<0$. 
We analyze this condition. 
By rewriting Eq.~(\ref{b}) to $R_0=R_\text{dS}/b$ and 
using the de Sitter solution $H_\text{dS}^2\equiv H_\text{dS+}^2$ 
with $\Lambda=\zeta/(\beta\kappa^2)$ in Eq.~(\ref{Hpm}), we get
\begin{equation}
0<\zeta<3\left(\frac{b-2}{b^2}\right)\,,
\quad 
0<\zeta<\frac{3}{8}\,,
\label{stab0}
\end{equation}
where we have also included the condition in Eq.~(\ref{Hpm}). 
For example, in the setting analyzed in the preceding subsection 
to reproduce the correct number of $e$-folds of inflation $\mathcal N$ in the model (\ref{model}), 
where $b=3$ and $\zeta=1/8$, 
this condition is met. {}From Eq.~(\ref{Delta0}), we can also estimate 
the time necessary for the universe to exit from inflation in the model (\ref{model}) as 
\begin{equation}
\Delta t\sim\frac{\pi}{2 B_0}=\frac{1}{H_\text{dS}}
\sqrt{\frac{b(6\Lambda_\text{eff}-\delta R_0^2\kappa^2)}{12(R_0-2\Lambda_\text{eff})}}\,.
\end{equation}
Since we are interested in the Planck energy, we may consider $c_0\sim M_\text{Pl}$ in Eq.~(\ref{Delta0}). 
For instance, we may be out from the Planck energy scale when $\Delta H\sim 10^{-5} c_0$, so that $\left|f(R)\right|\leq10^{-5} M_{\text{Pl}}^2$ and the effects of quantum anomaly disappear. 
On the other hand, the radiation, that we have neglected during inflation at the Planck scale as done in Eq.~(\ref{00}), becomes dominant and drives the cosmological evolution. 

We also consider the reheating stage after inflation. 
It is seen from the gravitational field equation that for 
an action in the Jordan frame such as that in Eq.~(\ref{action}) with 
Eq.~(\ref{model}), 
the Ricci scalar has a damped oscillating behavior~\cite{Mijic:1986iv}. 
In general, in the reheating process, the production of the particles happens 
in a similar way to that at the ordinary reheating stage 
in the Einstein frame. 
There are two cases where the action consists of 
a part of gravity and that of a scalar field. 
The first is the case of non-minimal 
gravitational coupling of a scalar field. This can be interpreted as 
the particle production from the gravitational effect in the regime of 
the perturbation~\cite{Mijic:1986iv, V-S}. The resultant energy density 
of the radiation produced at the reheating stage is known to proportional to the inverse of duration of the reheating stage. 
The second is the case that 
the particle production is realized by a coupling between the Ricci scalar 
and the square term of a scalar field~\cite{BD-MW} the value of which is larger than or equal to order of unity. This is called preheating through the parametric resonance phenomenon~\cite{PRH}. 

Furthermore, if we move to the Einstein frame from the Jordan frame 
with the conformal transformation, the reheating process due to the conformal 
anomaly can occur~\cite{Gorbunov:2012ns}. The conformal transformation leads 
to the coupling of the conformal scalar, called scalaron, with the trace of the energy momentum tensor of matter. Thanks to this coupling, the scalaron decays into the relativistic particles, and eventually the universe enters the radiation-dominated stage. The reheating temperature at the time when the energy density of the scalaron becomes equal to that of the radiation is proportional to 
$\sqrt{\Gamma_\mathrm{s}}\kappa$ with $\Gamma_\mathrm{s}$ the decay rate of the scalaron~\cite{Gorbunov:2010bn}.

\subsection{Effective gravitational coupling}

In order to avoid anti-gravity effects in a modified gravity theory, 
it is necessary to verify that 
$f_R (R)> -1$, namely, the effective gravitational coupling $G_{\mathrm{eff}}=G/\left(1+f_R(R)\right)$, where $G$ is the Newton's constant, have to be positive to correctly describe the interactions between matter and gravity. In the model (\ref{model}), $f_R(R)$ has a minimum at $R=0$ (this means the regime 
where $R/R_0\ll 1$), where
\begin{equation}
f_R(R=0)=-\frac{2\Lambda_\text{eff}}{R_0}\,.
\end{equation}
To recover the standard evolution of the universe for $R/R_0\ll 1$, 
we must have $G_\text{eff}\simeq G$. 
Moreover, when $R/R_0\ll 1$, by plugging Eq.~(\ref{Rll}) into the first Friedmann equation (\ref{EoM-00}), we obtain
\begin{equation}
3\Delta H(t)^2= \kappa^2\delta\left(6\Delta\ddot H(t) \Delta H(t)-3\Delta \dot H(t)^2\right)
+6\Lambda_\text{eff}
\left(
\frac{3\Delta\dot H(t)^2}{R_0^2}+\frac{\Delta H(t)^2}{R_0}-\frac{6\Delta H(t)\Delta\ddot H(t)}{R_0^2}
\right)+\kappa^2\frac{\rho_0}{a^4}\,,
\end{equation}
where we have taken account of the contribution of radiation. 
If $\Delta H(t)$ remains close to zero, that is, if Eq.~(\ref{RL}) is met, since $\kappa^2=8\pi/M_{\text{Pl}}^2$, the contribution of the quantum anomaly is completely negligible in comparison with that of radiation ($\kappa^2\rho_{\text{CA}}\ll 3\Delta H(t)^2$ with $\rho_{\text{CA}}$ the energy density of 
the conformal anomaly), but 
the contribution of the modification of gravity leads to
\begin{equation}
\kappa^2\rho_\text{MG}(R/R_0\ll 1)\simeq 3\Delta H(t)^2\left(\frac{2\Lambda_\text{eff}}{R_0}\right)\,. 
\end{equation} 
Therefore, only if $2\Lambda/R_0\ll 1$, the effects of modified gravity 
can be avoided. 
We conclude that in order to recover the correct value of the Newton's constant and the standard model for $R/R_0\ll 1$, the condition in Eq.~(\ref{RL}) must 
be well satisfied as 
\begin{equation}
\frac{2\Lambda}{R_0} \equiv
\frac{2\zeta b}{3\left(1+\sqrt{1-\frac{8\zeta}{3}}\right)}\ll 1\,.
\end{equation}

We have previously seen that the choice $\zeta=1/8$ and $b=3$ brings to a viable reproduction of the inflation in the model (\ref{model}). For this choice, we find $2\Lambda_\text{eff}/R_0\simeq 0.1$. 
Here, we present several comments. 
The smaller $\zeta$ is (that is, the smaller the effective cosmological constant of the exponential gravity model is), the better all the viability conditions of inflation are met. 
For example, if $\zeta=1/16$, i.e., 
$\Lambda_\text{eff}=\pi M_\text{Pl}^2/(2N^2)$, 
we acquire $2\Lambda_\text{eff}/R_0\simeq 0.06$. 
In addition, for $b=3$, the unstable de Sitter solution of inflation is realized, but the stable one is not realized. 
Furthermore,  
if $\gamma>4$, the number of $e$-folds $\mathcal{N}$ is larger than $76$. 
This fact is not surprising, because for $\Lambda_\text{eff}\rightarrow 0$, 
the modification of gravity disappears and we recover the Starobinsky model, 
which possesses an unique unstable de Sitter solution. 
However, in such a case, we also lose the effects of the perturbations from modified gravity, and an alternative description (like the usual one in scalar tensor theories) has to be found. 
To execute it is contrary to the aim of this work, in which we have demonstrated how it is possible to realize inflation (and in particular, the exit from inflation) in the framework of $F(R)$ gravity by taking into consideration the quantum anomaly combined with modified gravity. 

We finally state the following point. 
The toy model under investigation can easily be implemented in the form in 
Eq.~(\ref{proto2}) to unify the early-time inflation with the late-time acceleration
\begin{equation}
f(R)=-2\Lambda\left[1-\exp\left(-\frac{R}{R_*}\right)\right]
-2\Lambda_\text{eff}\left[1-\exp\left(-\frac{R}{R_0}\right)\right]\,,
\end{equation}
so that the modification of the Einstein gravity can be represented in a compact way to reproduce the accelerated expansion of the universe 
at two different energy scales. 
In the exponential model, there exists a stable de Sitter solution, but owing to the coupling with the trace anomaly at the Planck epoch, the solution at 
that time is unstable. 
As a result, the perturbations which make 
the exit from inflation towards the thermal universe at small curvature  
possible are originated from modified gravity itself.

\section{Unification of the trace-anomaly driven inflation with the dark energy dominated stage in exponential gravity} 

The first proposal on the unification of inflation and the dark energy 
dominated stage in $F(R)$ gravity has been made in Ref.~\cite{Nojiri:2003ft}. 
In this section, 
we show that actually, such a proposal works well also when 
the inflationary sector is represented by the trace-anomaly induced action. 
Indeed, this action is given by the specific form of modified gravity which 
contains an $R^2$ term and a higher-derivative non-local gravity one. 
Note that as it was demonstrated in Ref.~\cite{Bamba:2008ut}, 
such a term also removes possible early-time or late finite-time 
singularities by acting them towards a non-singular description of 
the universe. 

It follows from Eqs.~(\ref{action}) and (\ref{model}) that 
the action for the unified scenario is expressed as 
\begin{equation}
I=\frac{1}{2\kappa^2}\int_\mathcal{M} d^4 x \sqrt{-g}\,\left\{R
+\frac{\kappa^2}{96 \pi^2} \gamma N^2 R^2 
-2\Lambda_{\text{eff}}\left[1- \exp\left(-\frac{R}{R_0}\right)\right] 
+2\kappa^2\mathcal{L}_\mathrm{QC}\right\}\,. 
\label{eq:V1}
\end{equation}
Here, $R_0$ can be considered to be a current curvature. 
In this case, (i) for large curvature regime 
$R/R_0 \gg 1$, which corresponds to 
the early universe, the effective action can be written as 
\begin{equation}
I_\mathrm{inf} = \frac{1}{2\kappa^2}\int_\mathcal{M} d^4 x \sqrt{-g}\,\left(R
+\frac{\kappa^2}{96 \pi^2} \gamma N^2 R^2 
-2\Lambda_{\text{eff}} 
+2\kappa^2\mathcal{L}_\mathrm{QC}\right)\,. 
\label{eq:V2}
\end{equation}
While, (ii) for small curvature regime $R/R_0 \ll 1$, when the universe 
is at the dark energy dominated stage, the effective action can be 
represented as 
\begin{equation}
I_\mathrm{DE} = \frac{1}{2\kappa^2}\int_\mathcal{M} d^4 x \sqrt{-g}\,
\left(R-2 \tilde{\Lambda}_{\text{eff}} \right)\,, 
\quad 
\tilde{\Lambda}_{\text{eff}} \equiv \left(\frac{R}{R_0}\right) \Lambda_{\text{eff}}\,.
\label{eq:V3}
\end{equation}
In the expressions for the actions in Eqs.~(\ref{eq:V2}) and (\ref{eq:V3}), 
we only take the dominant terms. 
For the action in Eq.~(\ref{eq:V2}), thanks to the second term proportional 
to $R^2$, $R^2$ inflation can happen, and also the exit from inflation 
can be realized by the fourth term coming from the conformal anomaly. 
On the other hand, for the action in Eq.~(\ref{eq:V3}), 
the effective theory is equivalent to the Einstein gravity plus 
the positive cosmological constant, and therefore 
the de Sitter expansion of the current universe can occur. 
In this case, we have $a(t) = \bar{a} \exp\left(H_\mathrm{p} t \right)$ 
with $\bar{a} (> 0)$ a positive constant 
and $H_\mathrm{p} = \sqrt{\tilde{\Lambda}_{\text{eff}}/3}$ 
the Hubble parameter at the present time. 
As a result, in the model whose action is described by Eq.~(\ref{eq:V1}), 
not only inflation in the early universe 
but also the current cosmic accelerated expansion 
can happen in the unified manner as explained above.

\section{Conclusions and discussions}

In this paper, we have studied the conformal anomaly for inflation 
in $F(R)$ gravity. Since the discovery of the cosmic acceleration, the modified theories of gravity have become an interesting field of research. 
The idea that some modification of the Einstein gravity lies behind our universe offers a natural way to describe the accelerated expansion of the universe 
at large or small curvature. 
Inflation from $F(R)$ gravity is usually analyzed in the framework of the 
scalar-tensor theories, but we have here presented a different approach without invoking the conformal transformations for an $F(R)$ gravity model, which are 
even not always possible to do. Moreover, the conformal anomaly has been 
considered. If inflation takes place at the Planck epoch, the quantum effects 
must be taken into account. 

In our analysis, as a simple and suitable toy model, 
we have used exponential gravity with the quantum contribution due to the conformal anomaly. The accelerated expansion is initially driven by the conformal anomaly. Hence, the modification of gravity generates the perturbations necessary for the graceful exit from inflation. 

For our purpose, we have adopted the exponential gravity model, because in it, 
the de Sitter expansion is supported at the Planck epoch, and 
the negative perturbations of the curvature, 
which starts to decrease towards the end of inflation, are involved. 
It is well known that the conformal anomaly possesses the unstable de Sitter solution as an exact solution of the theory. This means that, given a perturbation, the model exits from the de Sitter solution. However, in order to understand 
the evolution of the universe and compare a model with the data obtained 
from the cosmological observations, it is crucial to reveal the origin and the form of the perturbations. In our case, the perturbations are generated from modified gravity itself and can be calculated. 
Consequently, the analysis of inflation can be carried out by analogy with scalar tensor theories. For our model, we can evaluate the number of $e$-folds $\mathcal{N}$, the spectral index, the time and the EoS parameter during inflation, and compare them with the latest observations. 
We have shown that the viable inflation occurs in the model, 
and at the end of inflation, the standard cosmology is recovered. 
Namely, the modification of gravity and the effects of the conformal anomaly vanish. Accordingly, radiation, which is, in principle, also contained in the quantum state, becomes dominant and drives the decelerated expansion. 

Concretely, it has been found that for our model, 
the spectral index of scalar modes of the curvature perturbations 
is $n_\mathrm{s} = 0.9603$, which is consistent with 
the Planck result, and 
the tensor-to-scalar ratio $r$ of them can be within the error of 
$68\%\,\mathrm{CL}$ of the very recent BICEP2 experiment data. 
It is also expected that in the near future, 
the B-mode polarizations of the CMB radiation might be 
detected by other experiments, e.g., 
QUIET~\cite{QUIET}, B-Pol~\cite{B-Pol}, 
POLARBEAR~\cite{POLARBEAR}, and LiteBIRD~\cite{LiteBIRD}. 

Finally, we remark that 
our approach can be generalized to other models of modified gravity. 
To realize inflation, it is required that modified gravity supports the de Sitter solution together with the conformal anomaly and generates the negative perturbations on the Hubble parameter.

\section*{Acknowledgments}

We would like to thank Prof. S.~Zerbini for useful discussions. 
K.B. is sincerely grateful to 
Prof. Gi-Chol Cho, Prof. Shin'ichi Nojiri, Prof. Akio Sugamoto
and Prof. Koichi Yamawaki for their very warm continuous encouragements. 
The work has been supported in part by the project TSPU-139 of Min. 
of Education and Science 
(Russia) (S.D.O.), and the JSPS Grant-in-Aid
for Young Scientists (B) \# 25800136 (K.B.).


\end{document}